# Invariant inter-subject relational structures in the human visual cortex

Ofer Lipman[1]  Shany Grossman[2,3,4]  Doron Friedman[5]  Yacov Hel-Or[1] and Rafael Malach[6*]

[1] Efi Arazi School of Computer Science, Reichman University, Herzliya, Israel
[2] Max Planck Institute for Human Development, Berlin, Germany
[3] Max Planck UCL Centre for Computational Psychiatry and Ageing Research, Berlin, Germany
[4] Institute of Psychology, Universitsät Hamburg, Germany
[5] Sammy Ofer School of Communications, Reichman University, Herzliya, Israel
[6] Department of Brain Sciences, Weizmann Institute of Science, Rehovot, Israel
* Corresponding author

**Abstract**

It is a fundamental of behavior that different individuals see the world in a largely similar manner. This is an essential basis for humans' ability to cooperate and communicate. However, what are the neuronal properties that underlie these inter-subject commonalities of our visual world? Finding out what aspects of neuronal coding remain invariant across individuals' brains will shed light not only on this fundamental question but will also point to the neuronal coding scheme at the basis of visual perception. Here we address this question by obtaining intracranial recordings from three cohorts of patients taking part in a different visual recognition task (overall 19 patients and 244 high-order visual contacts included in the analyses) and examining the neuronal coding scheme that was most consistent across individuals' visual cortex. Our results highlight relational *coding* – expressed by the set of similarity distances between profiles of pattern activations – as the most consistent representation across individuals. Alternative coding schemes, such as population vector coding or linear coding, failed to achieve similar inter-subject consistency. Our results thus support *relational coding* as the central neuronal code underlying individuals' shared perceptual content in the human brain.

**Introduction**

A fundamental aspect of our visual world, an aspect that is so basic that it is often taken for granted, is that different persons appear to see the world in a similar manner[1]. Such inter-subject agreement is essential for our ability to point to visual targets using language and to succeed in cooperative actions. The obvious challenge to visual neuroscience research is then to explain these inter-subject commonalities of visual perception in terms of cross-individual congruency in the neuronal codes underlying our visual experiences. Simply put, we can ask: what is kept similar in the neuronal coding of visual information as we move from one individual's brain to another? It should be noted that this question is relevant not only to explaining our shared visual world but also touches on the very basis of the neuronal coding mechanism that underlies the emergence of the rich array of our visual percepts.

We addressed this issue by studying neuronal responses in a large cohort of patients that underwent intra-cranial recordings in the course of a clinical diagnosis of epilepsy. This method provides highly detailed information, both spatially and temporally, of human neuronal responses. In the experiment, the patients were presented with the same set of pictures of different categories while their neuronal responses were recorded, allowing direct comparison of neuronal responses across patients. In the rest of the manuscript, we will term these functional congruencies between patients "inter-subject invariance".

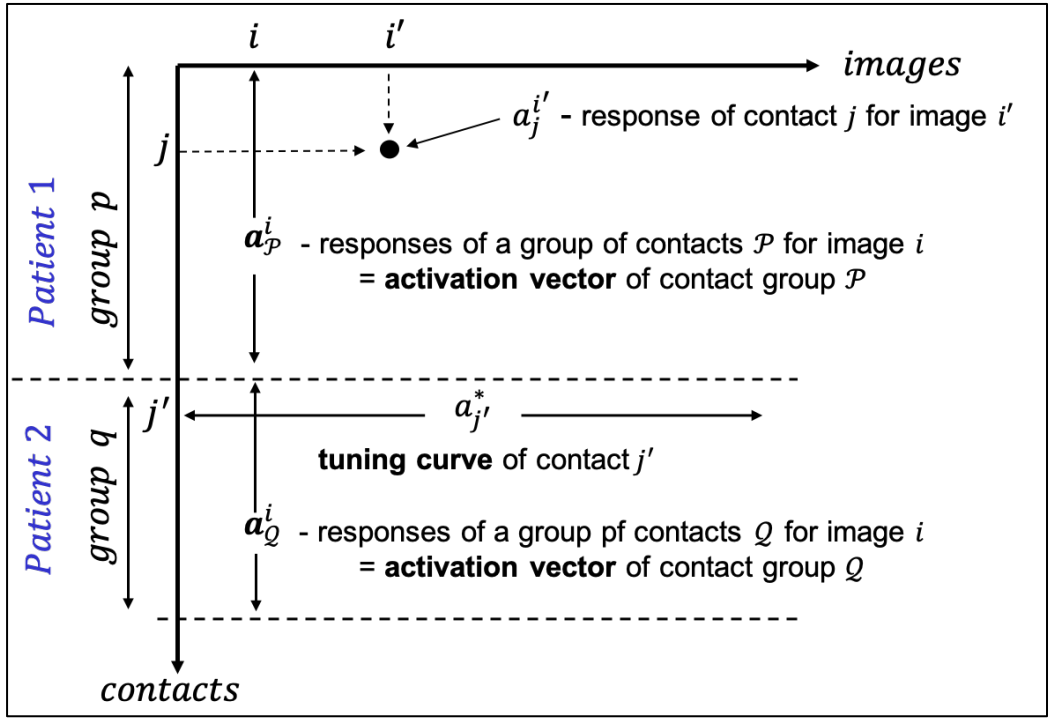

**Figure 1**. Population profiles of intra-cranial recordings captured as a matrix. Each row indicates a particular contact response vs. visual stimuli (images). Each column indicates the responses of all contacts for a particular stimulus. A group of contacts, $\mathcal{P}$ or $\mathcal{Q}$, indicates a set of contacts in a particular subject and in a particular cortical area. A set of responses for a particular group, representing a profile of pattern activation, is rearranged in a vectorial form denoted as an ***activation vector***. For example, the activation vector $\boldsymbol{a}_\mathcal{P}^i$ indicates the responses of all contacts in group $\mathcal{P}$ when the subject is viewing stimulus (image) $i$.

Formally, we denote by $a_j^i$ the activation of a contact $j$ when the subject is viewing an image $i$. Having a group of contacts, $\mathcal{P}$, of a particular subject in a particular cortical area, we define the population profile of these contacts into as an ***activation vector*** $\boldsymbol{a}_\mathcal{P}^i$ where the dimensionality of $\boldsymbol{a}_\mathcal{P}^i$ equals the number of contacts in the group $\mathcal{P}$ (see Figure 1).

Assuming two groups of contacts, $\mathcal{P}$ and $\mathcal{Q}$, from two different brains in the same cortical areas (let us assume for now that the two groups include the same number of contacts), we examined the level of inter-subject invariance under three hypothetical coding schemes, which are illustrated graphically in Figure 2.

The most straightforward scheme we considered is that perceptual information is encoded by the activation vector of cortical neurons. This coding scheme, denoted **activation pattern coding**, is arguably the most commonly considered in cortical neurophysiology [2–4] and is illustrated in Figure 2 panel (a). Here the assumption is that the direction of an activation vector elicited by a particular visual stimulus should be preserved across subjects. Using the above notations, for two brains with activation vectors $a_\mathcal{P}^i$ and $a_\mathcal{Q}^i$ respectively, we expect that $a_\mathcal{P}^i \propto a_\mathcal{Q}^i$, for each image $i$; namely, the activation vectors $a_\mathcal{P}^i$ and $a_\mathcal{Q}^i$ should be identical or proportional to each other, for each image $i$. In other words, we say that the two activation vectors are *directional invariant*.

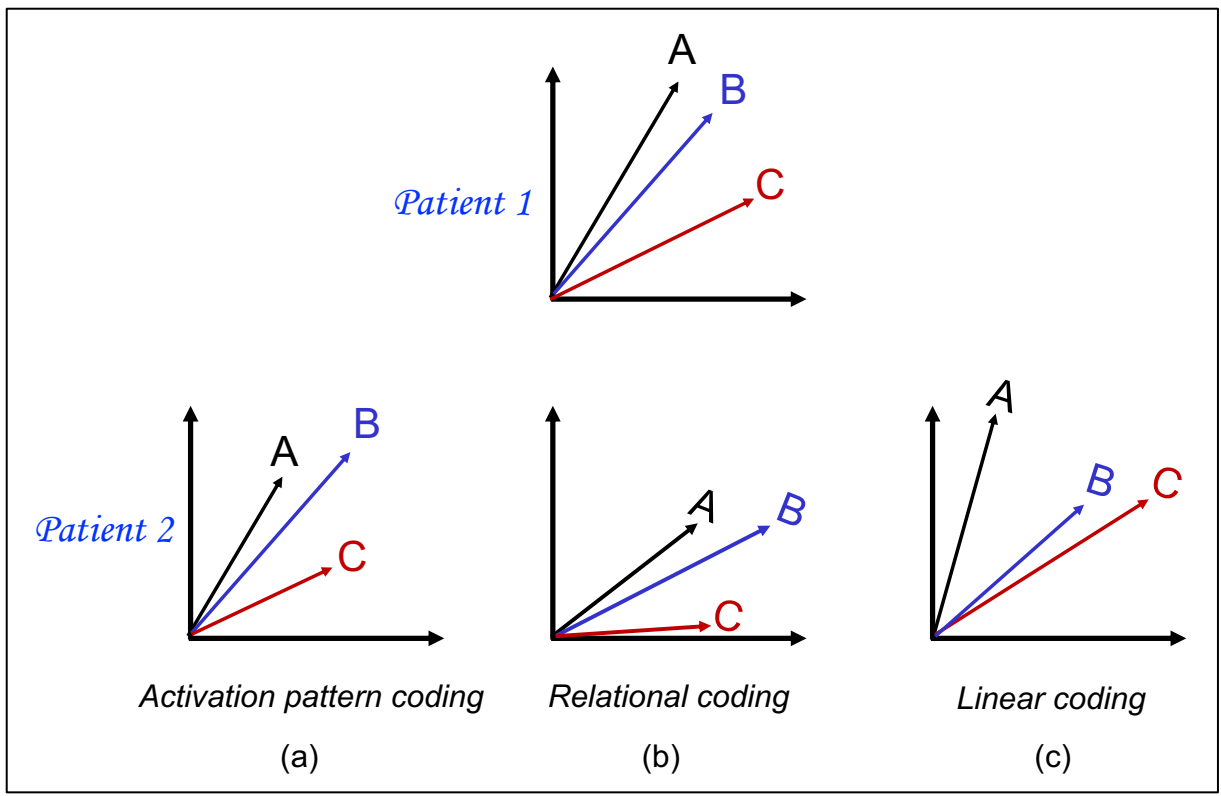

**Figure 2**. A schematic illustration of the three possible encoding schemes across subjects examined. (**a**) **Activation pattern coding,** retaining the orientation of activation vectors between subjects. (**b**) **Relational coding**, retaining the intra-angle distances between activation vectors, and (**c**) **Linear coding**, retaining the linear relations between triplets of activation vectors (see text). Axes x,y represent activities in contacts 1 and in contact 2.

We have previously presented data compatible with this very basic scheme, showing, using fMRI, an inter-subject congruency in the tuning properties of cortical sites during naturalistic visual stimulation[5,6]. However, it should be emphasized that due to the limited spatio-temporal resolution of fMRI, these findings are only relevant to the coarser aspects of the cortical visual representations – i.e., regions selective to general visual categories and entire cortical areas. Going beyond these coarse measures necessitates brain recording methods of higher spatio-temporal resolution; hence our choice to use direct intracranial recordings in patients[7–9].

An important point that needs emphasizing is that, as illustrated in Figure 2, panel (a), the *activation coding* hypothesis does not necessarily predict similar overall activations across individuals; only that their activation profiles – represented by their directions in the vector space – should be preserved. Thus, the "distance" between two activation vectors $a_\mathcal{P}$ and $a_\mathcal{Q}$ is defined by their "cosine distance": $dist(a_\mathcal{P}, a_\mathcal{Q}) = 1 - (a_\mathcal{P} \cdot a_\mathcal{Q})/(|a_\mathcal{P}||a_\mathcal{Q}|)$ where $a_\mathcal{P} \cdot a_\mathcal{Q}$ denotes an inner product between vectors, and $|a_p|$ indicates the L$_2$ norm. The assumption that the overall magnitude of activation is not a critical parameter in establishing perceptual content has recently received experimental support in explaining the perceptual stability of stationary images despite drastic changes in overall activation magnitudes[4,10]. Consequently, in the *activation pattern coding*, where we assume directional invariance, we expect that the $dist(a_\mathcal{P}^i, a_\mathcal{Q}^i)$ should be close to zero, for each image *i*.

While *activation pattern* has been considered the most basic cortical coding scheme, in recent years there has been a growing interest in the possibility of an alternative coding scheme, denoted **relational coding**, echoing classical structuralist principles[11,12]. Relational coding is a theory that proposes that the brain represents information by the *correlations between the activation vectors* elicited by different *objects* or concepts, rather than by the actual activation vectors themselves. This theory suggests that the brain encodes information about the world by creating representations of the relationships between objects and concepts, rather than representing each object or concept individually. In this model, visual information is not encoded by the activation vector on its own, but rather by the set of similarities and dissimilarities – also termed distances (see[4,12–15]) – that appear between the activation vectors when activated by different visual contents [12–16]

In term of the vector space of activation vectors, with relational coding we would expect different individuals to preserve the mutual distances between stimuli and not necessarily to preserve the original orientations of the activation vectors. Formally speaking, we say that activation vectors of two individuals are similar *up to rotation* (orthogonal transformation in high dimensional space, which is an angle preserving transformation) while relaxing the inter subject congruency between the activation vectors themselves. Using the above notations, it gives that for two brains with activation vectors $a_\mathcal{P}^i$ and $a_\mathcal{Q}^i$ respectively, we expect that there is an orthogonal transformation $R$, such that $Ra_\mathcal{P}^i \approx a_\mathcal{Q}^i$, for each image *i.* Namely, the activation vectors $a_p^i$ and $a_q^i$ are identical up to orthogonal transformation $R$. This yields that the angles between activation vectors in one subject are preserved in other subjects: $dist(a_P^i, a_P^j) = dist(a_Q^i, a_Q^j)$ for each pair of stimuli $i, j$. This invariance is a direct outcome of the property that orthogonal transformation is angle preserving. This prediction is illustrated in Figure 2 panel (b) termed *relational coding*. Some support for the notion that this indeed may be the invariant coding across individuals has been recently obtained using fMRI. Thus, [13,14] have shown that aligning the inter-pattern distances provides high consistency across individual brains, which is transferable across different visual stimuli.

Finally, we may continue along this line and increase the model complexity to the next step assuming that the perceptual code is invariant up to general linear transformation between individuals, namely, there is a general matrix M, such that $Ma_P^i \approx a_Q^i$ for each image $i$. Intuitively, a linear transformation is a rule that takes input vectors, and by stretching, shrinking, rotating, or flipping them, outputs new vectors (while straight grid lines before the transformation are kept straight after the transformation). This model is illustrated in Figure 2 Panel (c). In that case, we may expect the angle distances between activation vectors to vary between individuals – while preserving a fixed, linear transformation between them.

To explore the above three alternative predictions, we have assembled a large cohort (19 patients implanted with 244 visually responsive contacts) of intracranial recordings in patients with the aim of examining which of the three coding models is best conserved across individual brains.

**Results**

To examine inter-subject similarity of visual responses, patients undergoing intra-cranial recordings for clinical purposes (see Fig. 3 and Methods), were presented with identical sets of visual images of different categories – such as, faces, houses and tools (for more details see Methods), while their neuronal responses, measured as changes in high frequency broad band power of the iEEG signal (see Methods) were recorded. Our experiment is based on an initial cohort of three datasets including 68 patients and 852 visually responsive contacts. However, here, and for the rest of the paper, we confined the analysis to the high-order contact group (19 patients, 244 contacts, see Methods for inclusion criteria), suggested by a large body of prior research to be the central hub of perceptual content[12].

In the experiment (Fig. 3), visual Images were presented for 250 ms (sets 1 and 2) or 500 ms (set 3) interspersed with 500-750 1050 ms fixations. The patients' task was a 1-back recognition task, in which patients had to indicate rare images (12% for sets 1 and 2, 10% for set 3) repeats via a button press. The image repeats (i.e., the 1-backs) were discarded from all analyses.

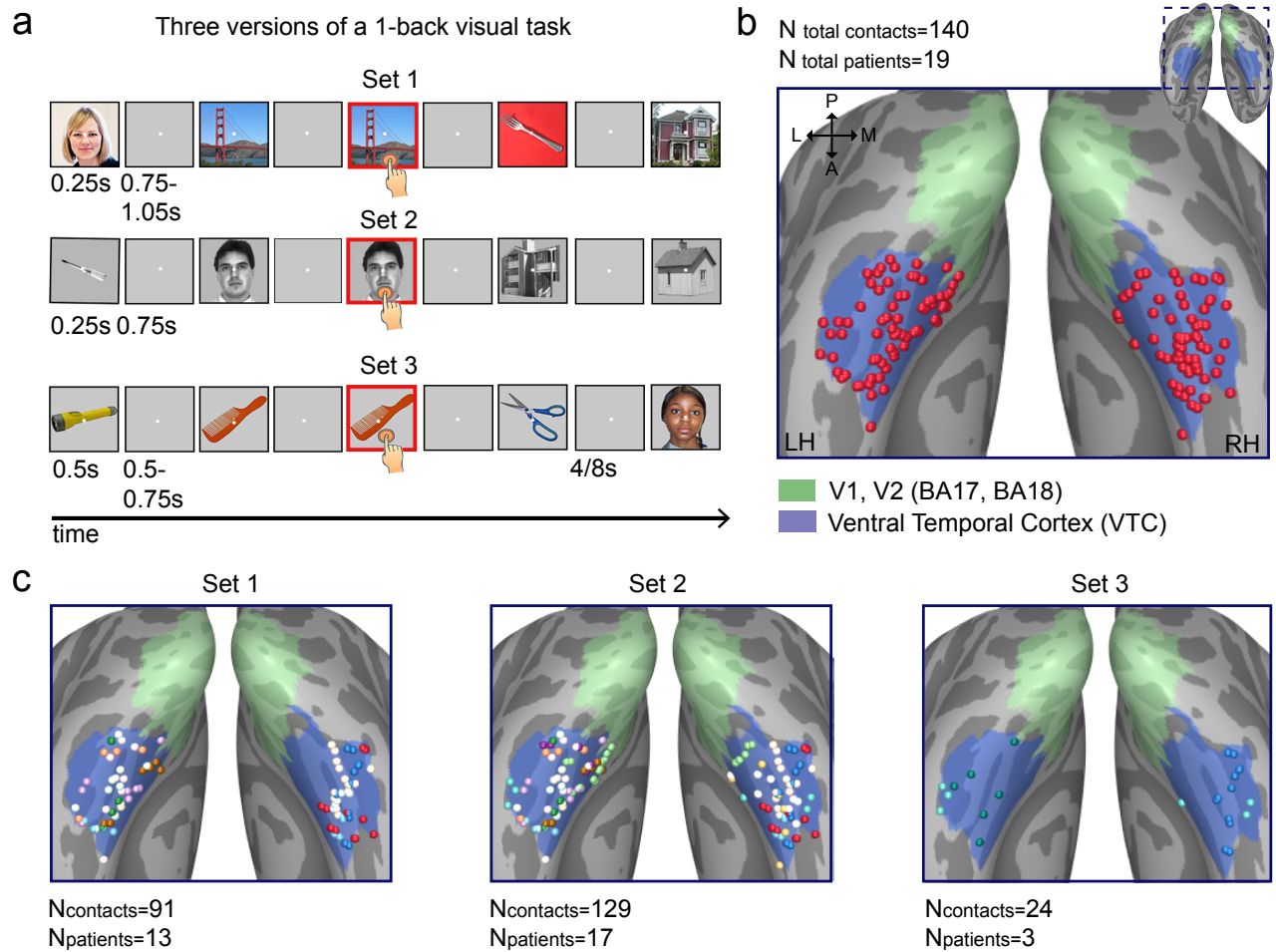

**Figure 3**. Experimental procedure and contacts localization. (**a**) All studies followed the 1-back visual recognition task in which patients were asked to press a button on rare occasions of image repeats. (**b**) Locations of contacts included in the analyses pooled across all 3 sets (red circles), after being projected onto a common cortical surface template (FSAverage). High-order (VTC) and low-order (V1,V2) visual areas are denoted by green and blue colors. (**c**) Contacts localization presented separately for each of the 3 sets. Contacts originating from different patients are denoted by different colors. Note that 14 patients took part in 2 out of the 3 tasks (therefore the total number of unique patients is smaller their sum across the 3 sets).

**The inter-subject similarity of anatomically neighboring contacts**

The simplest model of inter-subject similarity assumes that contacts that are located in close homologous anatomical locations across patients should also show similar functional tuning compared to contacts that are located anatomically in disparate locations. To examine if this indeed was the case in our data, we compared the activation similarity of all possible pairs of contacts and across all patient pairs in high order visual cortex. In our analysis, we calculated the inter-subject anatomical proximity of each pair of contacts – first by normalizing all patients' brains to a standard brain (see Methods). We then checked whether anatomical proximity is indeed translated to a more similar tuning curves for all pairs of contacts and across all individuals. The result of this large-scale analysis is depicted in Figure 4. In line with our previous fMRI study (e.g.,[5]) – contact pairs that were located in close anatomical positions showed a significantly higher degree of functional correlation compared to distant pairs (p=0.04) (see Methods for more details). However, although significant, the correlation value was extremely low, revealing a high degree of local scattering, which clearly violated the anatomical proximity rule.

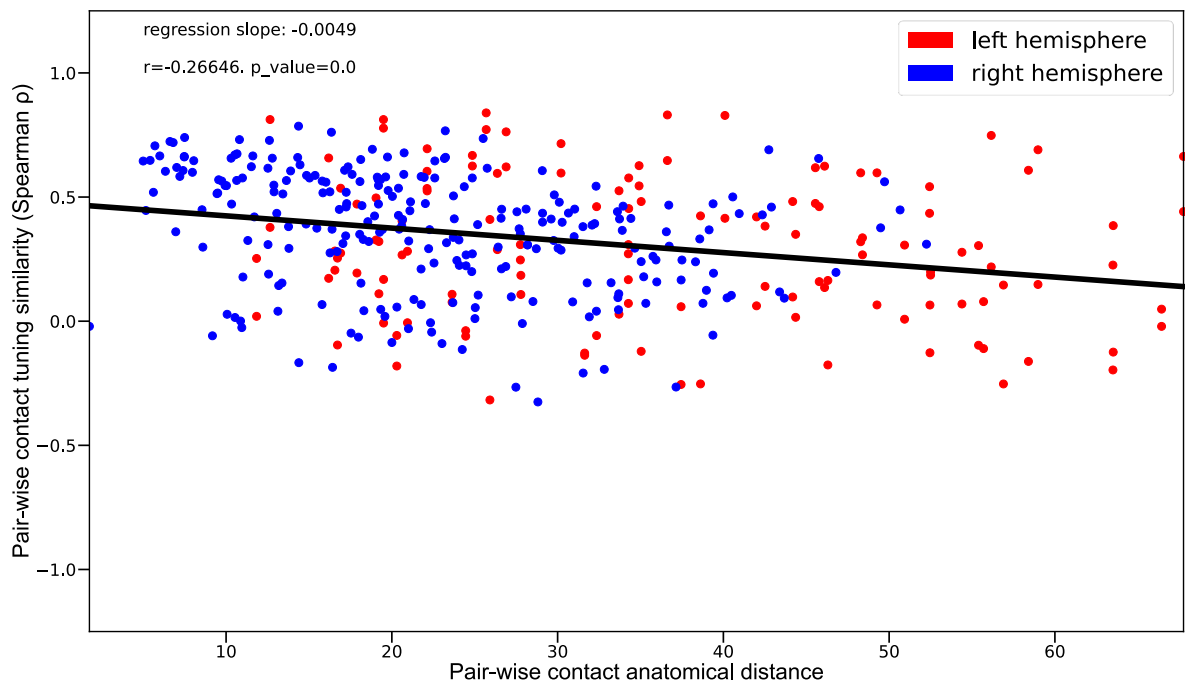

**Figure 4**. Functional similarity between pairs of contacts from different patients (19 patients, 244 contacts) as a function of their anatomical distances. The Y-axis depicts the similarity of the tuning curve measured by Spearman correlation between contact pairs, while the X-axis depicts their anatomical distance. Only contact pairs localized in the same hemispheres and in high order visual areas were evaluated. Note the significant but extremely low correlation, indicating a high degree of local scatter in functional similarity across patients.

**Pairing contacts between subjects**

Nevertheless, it could be argued that although homologous cortical locations may not be linked to identical tuning properties of individual contacts, such similarities in functional tunings may be apparent at the levels of groups of contacts, regardless of their precise location within the group, i.e., across activation patterns (see definition in Figure 1) rather than individually matched homologous contacts. According to the hypothesis presented as "activation vector coding" in Figure 2, we do not confine the search to exactly matched anatomical locations across individuals, but rather evaluate the similarity between contacts based on their corresponding activations.

In order to perform this location-free analysis, we derived an optimal matching between contacts across each pair of individuals in the following manner: For a specific contact, say contact $j$, we define the *tuning curves* $a_j^*$ as the contact values across all images (see Figure 1). To calculate the similarity between two tuning curves across two contacts, $a_j^*$ and $a_k^*$, we take the Spearman's rank correlations between $a_j^*$ and $a_k^*$, denoted by $S(a_j^*, a_k^*)$. For two groups of contacts, $\mathcal{P}$ and $\mathcal{Q}$, each of which belongs to a different patient, we define the similarity score as the mean of the Spearman's correlation:

$$S_{\mathcal{P},\mathcal{Q}} = \frac{1}{n} \sum_{j \in \mathcal{P}} S(a_j^*, a_{j'}^*)$$

where for each contact $j$ in group $\mathcal{P}$, we denote $j' = \Pi_\mathcal{Q}(j)$ as the matched contact in group $\mathcal{Q}$. We assume we have *n* contacts in each group, thus we normalize by *n*. The best match of the two groups can be sought such that the similarity score will be maximal:

$$\Pi^* = argmax_\Pi \sum_{j \in \mathcal{P}} S(a_j^*, a_{j'}^*) \quad \text{where} \quad j' = \Pi_\mathcal{Q}(j)$$

This optimization can be formalized as a bipartite graph matching problem[15] that finds the optimal matching between contacts in group $\mathcal{P}$ (patient A) and contacts in group $\mathcal{Q}$ (patient B). If the number of contacts in the two brains is not identical, the solution provides the maximum number of possible pairs (that is, the contacts' count in the patient with the smaller number of contacts).

**Inter-subject Activation pattern Invariance**

In order to compare the three hypotheses, we define a measurement between subjects that we term Inter-Subject Similarity Index (ISSI). For the first hypothesis, as outlined above, instead of selecting homologous cortical locations, we search for the best possible one-to-one mapping of contacts between each pair of subjects. Specifically, through permuting all possible combinations of contacts between each pair of patients, the maximum overall Spearman correlation over all possible pairs of patients was established (see Methods). We then calculate the ISSI scores for the activation vectors across all images as follows: Assuming a pair of patients with corresponding sets of contacts, $\mathcal{P}$ and $\mathcal{Q}$, where the permutation matrix $P$ is the optimal permutation between $\mathcal{P}$ and $\mathcal{Q}$ according to the best matching procedure (the matrix $P$ is extracted from the permutation mapping $\Pi$ above). The ISSI between $\mathcal{P}$ and $\mathcal{Q}$ is calculated based on the Spearman correlation:

$$ISSI(\mathcal{P},\mathcal{Q}) = \sum_i S\left(P\boldsymbol{a}_\mathcal{P}^i, \boldsymbol{a}_\mathcal{Q}^i\right)$$

where in this case, the correlations are summed over a set of images (index $i$), and the Spearman correlation is calculated between activation vectors rather than tuning curves.

Critically, this analysis was performed on a randomly selected subset of images, that will be termed the **training set** (80% of the data). Once the contact set was selected based on this training set, the ISSI was calculated on the remaining images that were not part of the optimization process, which will be termed the **validation set**.

The results of this analysis of the activation are presented in Figures 5 (red curves) and in Figure 6 (red bars). Figure 5 shows an example of ISSI values of the activation vectors across all pairs of patients, when they were presented with the category of animal images in set 1. We plot all ISSI values in an increasing order. The lower panel shows the ISSI for the training set of images, while the upper panel shows the ISSI values for the validation set. As can be seen, there was a small positive and consistent similarity across individuals (bottom panel) that – importantly – was maintained in the independently measured validation set (top panel). Supplementary Figure 1 depicts the same analysis for the entire an additional set of image categories showing, again, a small but consistent increase in the ISSI values of the activation vectors across all object categories. Figure 6 depicts, for each category, the average ISSI values across all patient pairs for the activation vector similarity analysis (red bars). Similar to the results of Figure 5, we see a small and positive inter-patient similarity in this measure across all different categories, both in the training as well as the validation sets.

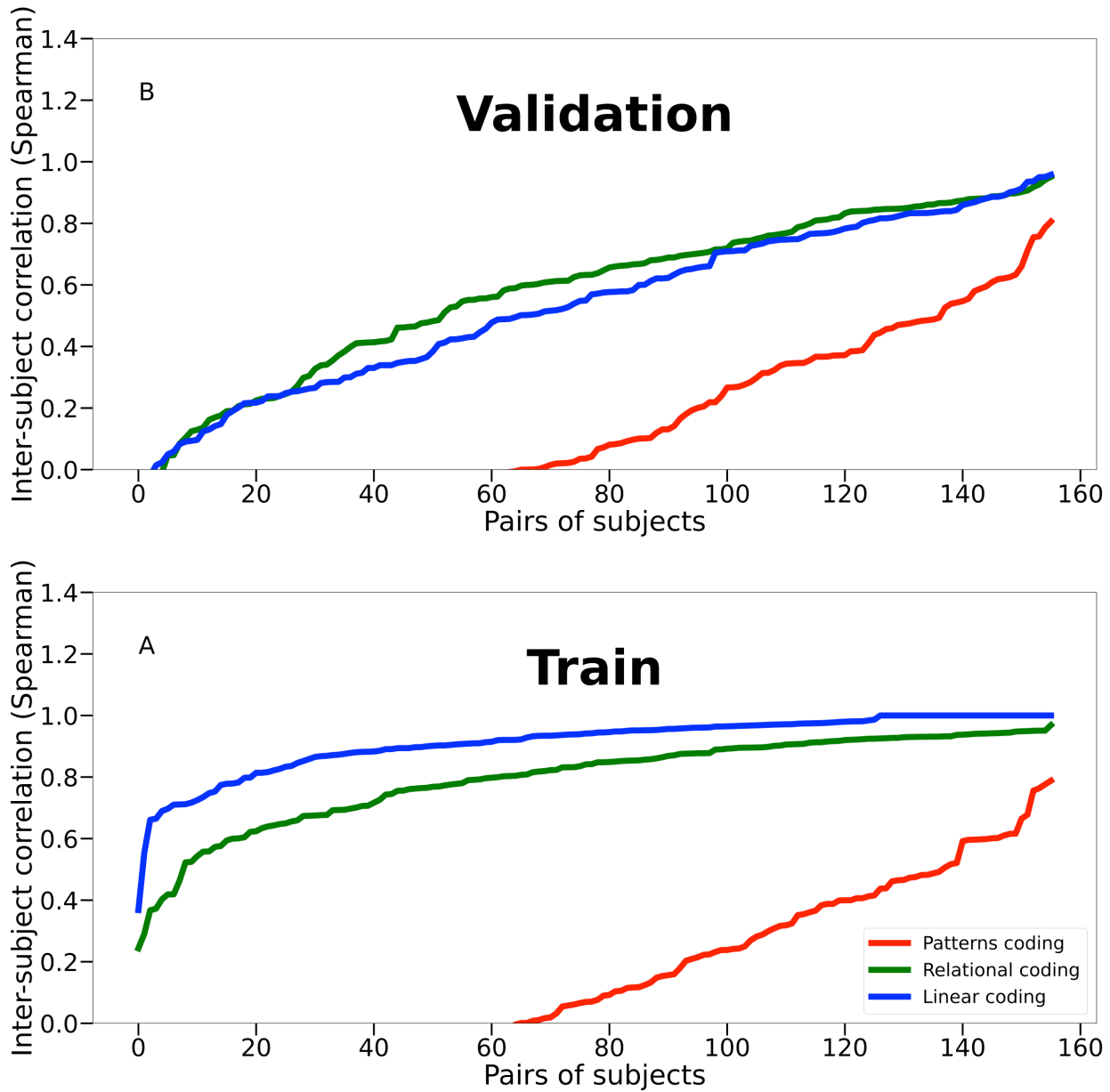

**Figure 5**: Inter-patient ISSI of the three neuronal coding schemes for the animals category (set1). The Y-axis depicts the Spearman correlation between pairs of patients. X-axis depicts the pair index. The ISSI values are sorted and shown in increasing order. Bottom panel: training set. Top panel: independent validation set. Red, green, and blue colors indicate the ISSI for optimal permutation (activation vector matching), rotation (relational code matching), and linear transformations matching, respectively. All three coding schemes show a significant measure of inter-subject invariance; however, for the validation set, the relational coding scheme (rotation) shows a clear enhancement compared to the other coding schemes.

**Inter-subject relational coding Invariance**

The second hypothesis we consider is that of *relational coding*. In this scheme, the neural representation of a concept is based on how similar and dissimilar this concept is from other encoded concepts. Formally, a *relational coding scheme* implies that the activation codes are invariant across individuals up to orthogonal transformations, namely, there is an orthogonal transformation $R$, such that, for each image $i$, $R\boldsymbol{a}_\mathcal{P}^i \approx \boldsymbol{a}_\mathcal{Q}^i$. The orthogonal transformation, $R$, which can be viewed as rotation in high-dimensional space, is a distance-preserving transformation, thus, if indeed the coding between individuals is identical up to rotation, we have:

$$dist(\boldsymbol{a}_\mathcal{P}^i, \boldsymbol{a}_\mathcal{P}^j) = dist(\boldsymbol{a}_\mathcal{Q}^i, \boldsymbol{a}_\mathcal{Q}^j) \quad for\ each\ pair\ of\ images\ (i,j)$$

Here, the activation vectors are not preserved across individuals, but rather the inter (angle) distances between these vectors. At this point, it is worth mentioning that invariance up to permutation (the 1st hypothesis) is a special case of invariance up to rotation (permutation is a particular case of rotation). Therefore, it is expected that the ISSI score after optimal rotation will exceed the ISSI after optimal permutation (1st coding scheme). However, this will not necessarily be the case when the rotation is determined on a training set and evaluated on an independent, different, validation set.

To examine whether indeed such relational coding was preserved across patients we conducted the following analysis. First, for each category of visual stimuli and each pair of patients we applied the optimal-match procedure as discussed above. After this step, in each pair of patients the same number of contacts that are optimally correlated were selected for further analysis. Next, we divided the set of all visual stimuli into training and validation sets (on average, over 100 train/test splits, with each split 2/3 training and 1/3 validation). Next, we computed the optimal rotation (orthogonal transformation) for each pair of patients on the training set, and then applied the evaluated rotation to the validation set (see Methods). The optimal rotation, $\hat{R}$, was evaluated by solving (see Methods):

$$\hat{R} = argmin_R \sum_{i \in train\_set} \|R\boldsymbol{a}_\mathcal{P}^i - \boldsymbol{a}_\mathcal{Q}^i\|_2$$

To evaluate the similarity between the two patients after applying the optimal transformation we calculated the ISSI measure (as it is more robust for monotonic value mapping) again on the validation set:

$$ISSI(\mathcal{P}, \mathcal{Q}) = \sum_{i \in val\_set} S(\hat{R}\boldsymbol{a}_\mathcal{P}^i, \boldsymbol{a}_\mathcal{Q}^i)$$

The results of this analysis are presented by green curves in Figures 5 (lower and upper panels for training and validation set) for the category of animal images (and for all other categories in suppl Figures 1), and by green bars in Figure 6 for all categories. As can be seen, there was a robust and consistent increase in the inter-subject similarities when relational coding formed the basis of comparison as opposed to the activation pattern coding scheme. Such increase is indeed to be expected for the training set, however, critically, a robust and significant increase was also evident when examining the validation set (Wilcoxon signed-rank test between ISSI values for relational match and activation pattern match). The effect was consistent across all image categories (see Figures 5, 6 and suppl Figure 1).

Thus, we find that relational coding, as reflected in the rotational analysis, shows consistently and robustly, a better match across subjects and image categories compared to the activation pattern coding.

**Inter-subject Linear Invariance**

The observation that relational coding is matched across individual patients raises the possibility that perhaps this similarity will be further enhanced if we further relax the constraints on the image coding schemes. A straightforward next step in such a relaxation process can be applied by extending the transformation from orthogonal to linear transformations (see illustrations in Figure 2); as mentioned, if the orthogonal transformation yields representation invariance up to rotation, the linear hypothesis yields invariant representations up to a linear transformation. Such a transformation maintains ratios of distances between three co-linear points, but its rotation does not preserve distances and angles.

To examine this hypothesis, we conducted the same analysis as the one carried out for relational coding, only this time using linear transformations rather than orthogonal transformations. Note, also in this case, that the orthogonal transformation is a particular case of the more general linear transformation model. The optimal rotation of a linear transformation is evaluated by solving (for more details see Methods):

$$\widehat{M} = argmin_M \sum_{i \in train\_set} \left\| M \boldsymbol{a}_{\mathcal{P}}^i - \boldsymbol{a}_{\mathcal{Q}}^i \right\|_2$$

Where M is a general linear transformation matrix.

Evaluating the similarity between the two patients after applying the linear transformation is calculated, similarly, as follows:

$$ISSI(\mathcal{P}, \mathcal{Q}) = \sum_{i \in testval\_set} S(\widehat{M} \boldsymbol{a}_{\mathcal{P}}^i, \boldsymbol{a}_{\mathcal{Q}}^i)$$

The results for this transformation are presented in Figure 5 in blue curves (see suppl Figure 1 for all additional categories of images). As can be seen, in the training set, this richer transformation produced a higher level of similarities compared to the relational code. This is to be expected since the linear transformation has more degrees of freedom. However, critically, when examining the validation set, in most cases the similarities fell below the orthogonal transformation (non-parametric statistics on all categories, 9 out of 12 categories show a preference for relational over linear coding, p=0.05). This is further demonstrated in Figure 5 and suppl Figure 1. These results indicates that the enhanced power of the linear transformation likely resulted in over-fitting.

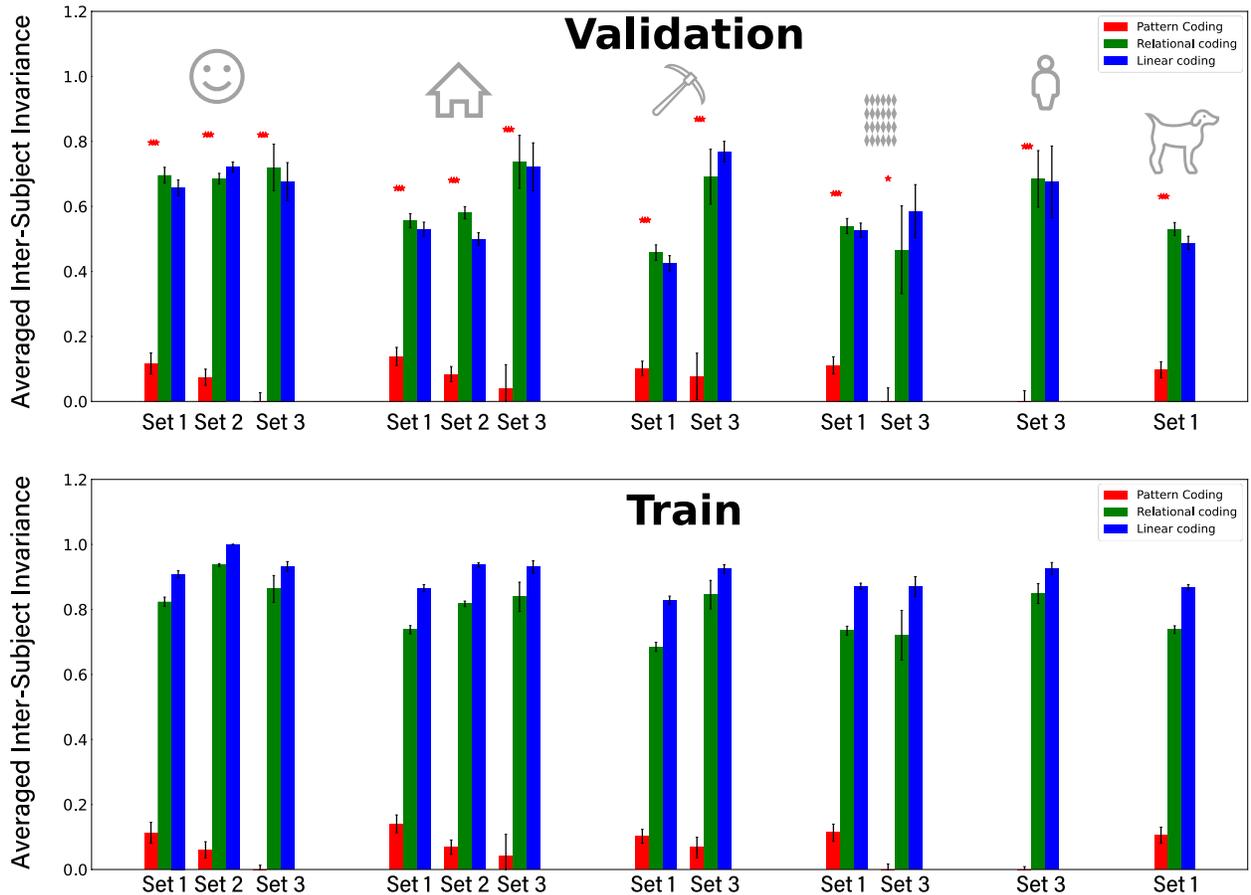

**Figure 6**: Comparing the averaged inter-subject invariance (ISSC) across the three coding schemes. Red, green, and blue bars depicting the ISSC for the activation pattern coding, relational coding and linear coding, respectively, across all subjects and image categories. Image categories from left to right: faces, houses, tools, patterns body and animals. The last set on the right is the averaged mean across all categories. Bottom set – training image set, top set – independently verified validation test. Note the striking and highly significant (wilcoxon signed-rank test) increase for the inter-subject invariance of the relational coding relative to the activation pattern coding. The small but significant (non-parametric statistics on all signed differences, relational higher than activation pattern in 9/12 samples, p=0.05) reduction in the linear transformation in the validation set suggests that the linear transformation overfits the inter-subject invariance (see Figure 1), leaving the relational coding scheme the best model for invariance.

**Neuronal coding across image categories**

To examine how consistent were these results across patient groups and image categories, Figure 6 compares the ISSI levels of the three neuronal coding schemes, calculated separately for the different image categories and for the contacts corresponding to the three independent datasets in high level visual areas (see Figure 3C). The lower panel shows the results for the training set while the upper panel show results for the validation data. As can be seen, and similar to Figure 5 and suppl Figure 1, there was a small and fairly consistent level of ISSI of activation vectors across patients as revealed by the positive red bars across most (9 out of 12) image categories. Importantly, this inter-subject activation pattern similarity was apparent also in the independently validated set. However, critically, there was a consistent and striking enhancement in the ISSI level when shifting to the relational coding scheme. The effect was consistent across all image categories (see green colors in Figure 5, Suppl Figure 1, and Figure 6) attesting to the cross-category superiority of relational coding.

## Discussion

**Inter-subject invariance as a window into cortical perceptual codes**

Our strategy in the current work was to search for coding schemes that are preserved across the visual systems of different individuals. As discussed in the introduction, finding such cross-subject invariances is of great interest since our visual world is shared to a large extent, and individuals most often agree on how things look like. It should be emphasized that, given the shared nature of the visual world, finding neuronal codes that are not preserved across individuals can be taken, tentatively, as an argument against their essential role in perception. By contrast, those codes that evolution appears to have preserved across individuals' visual systems are likely to play an essential role in visual perception. As we will discuss in detail below, following this logic and comparing three neuronal schemes of such perceptual coding clearly highlights *relational geometries* as an essential coding scheme of perceptual images in the human brain.

**Relational coding of visual perceptual content**

Our results reveal a weak but consistent inter-subject invariance in activation pattern coding (Figure 6). However, our central finding is that when activation vector for each image is allowed to vary across individuals – while the inter-vector angles are kept invariant, i.e., preserving the relational coding across individuals, the inter-subject similarities were strikingly and significantly enhanced (compare green vs. red lines and bars in Figures 5, 6 and suppl Figure 1).

It is important to clarify that it is to be expected that relational coding should be better matched across individuals compared to activation pattern coding when examining the training set alone. This may be a trivial consequence of the larger flexibility (more degrees of freedom) we allow in the comparison across patients in the former compared to the latter models. However, critically, the fact that such enhanced inter-subject similarity was transferred to the validation set – demonstrates that the preservation of relational similarity distances is a genuine cortical feature and not a technical consequence of a better fit due to a more flexible model. Moreover, following the same logic it is expected that the linear coding scheme should better match across individuals in the training set compared to the relational coding scheme, as orthogonal transformation is a special case of a linear transformation. However, it is apparent that in the validation set the relational coding still surpasses the linear coding although it has fewer degrees of freedom. Thus, our central finding points to relational coding as the essential representational scheme underlying the content of human shared perception.

It should be clarified in what manner relational coding is fundamentally different from the better-known neuronal tuning curves and activation vectors coding schemes. The fundamental difference is that according to the single unit and activation-vector schemes, the information pertaining to the visual stimuli is coded in the unique patterns of activations to each stimulus. By contrast, in relational coding, the specific population vector profile elicited by each stimulus is irrelevant to the perceptual content. Instead, what defines the content of a visual percept is the set of similarities and dissimilarities, i.e., the distances between the neuronal population profiles elicited by this percept and those of all other percepts. The neuronal mechanism underlying these relational geometries is embedded in the unique synaptic structure of each cortical region and is likely revealed through recurrent neuronal activations within each neuronal structure[12].

In recent years, relational geometries have emerged as a central concept in artificial networks, as well as in characterizing the primate visual cortex [16,17][18]. It may be informative to note here similar relational geometries found across different artificial networks[19]. Importantly, a growing number of studies have demonstrated a link between relational coding and behavior. These studies include perceptual similarity measures[20,21] semantic categories[22] emotional representations[23] and even in episodic memory semantization processes[24]. The present study extends these findings by demonstrating that despite the unique nature of individual visual representations, there is a cross-individual preservation of the relational geometries – thus highlighting this coding scheme as essential for human perception.

It is also interesting to note here that a similar invariance has been found in artificial neural networks (ANNs). Empirical evidence suggests that two corresponding layers in two different ANNs, when trained on the same dataset but with different initializations, exhibit similar responses up to orthogonal transformations[19]. This phenomenon corresponds to relational coding.

**Weak inter-subject tuning similarity across anatomically similar homologous loci**

Our results show a weak but significant similarity in tuning of individual contacts occupying similar anatomical locations across individuals (Figure 4). This result is compatible with previous findings of such inter-subject synchronization revealed through fMRI (e.g., Hasson et al., 2004). However, this similarity was extremely small at fine resolution, likely masked by a substantial level of local scatter (see Figure 4). A likely source for the weak relationship may be individual variabilities in the precise location of homologue cortical regions, which may occupy different anatomical loci relative to the major brain landmarks used for cortical alignments. Finally, it is plausible that the anatomical mosaic of functional properties at high spatial resolutions varies randomly across individual brains, while at a coarser level it is preserved. This will result in a lack of precise functional matching between homologous cortical points at the local scales within single cortical regions while showing consistent inter-subject correlations at coarser scales as was indeed demonstrated by [5].

**In conclusion**

We demonstrate here, using three iEEG data sets, that the neuronal perceptual coding scheme that is most preserved across individuals is *relational coding*. In this coding scheme, the similarity between activation vectors is maintained across individuals, rather than the tuning profiles of individual contacts or the activation vectors themselves. These findings offer a robust explanation for the ability of different individuals to perceive the world in a similar manner and point to *relational coding* as an essential cortical code underlying visual perception.

**Methods**

*Participants*

Data were retrieved from a study consisting of three tasks and reported in two previous studies[26,27]. Sixty-eight participants were monitored for pre-surgical evaluation of epileptic foci. Among these subjects, 19 had a minimum of 4 visually responsive selective contacts located in high-order visual areas (14 of which took part in two out of the three tasks, as detailed in Supplementary Table S1). All the participants gave their fully informed consent, including consent to publish, in compliance with NIH guidelines. This was done under the supervision of the institutional review board at the Feinstein Institute for Medical Research, in line with the principles of the Declaration of Helsinki.

*Experimental Design*

The three tasks performed were 1-back visual tasks. Each consisted of a different set of 10 face s images, houses/places, objects, patterns (set 1 and 3 only), animals (set 1 only) and bodies (set 3 only) and a different set of additional images from other categories (see Fig. 3a for a schematic illustration of the three tasks). Each patient performed either one or two versions of the three (for individual specifications see Supplementary Table 1).

The first task (set 1) consisted of a total of 60 stimuli: faces (public figures collected from the internet), words, animals, tools, patterns and places, 10 instances for each class. Throughout the task, each image exemplar was presented a total of six times. During the task, images were displayed for 250 milliseconds, followed by a jittered inter-stimulus interval ranging from 750 to 1050 milliseconds. The task consisted of 360 trials, including 24 1-back repetitions. The second task consisted of a total of 56 grey scaled stimuli: faces, tools, patterns, houses, and body parts, 10 instances for each class. During the task, each image exemplar was shown 3-4 times. The images were displayed for a duration of 250 milliseconds, at a constant frequency of 1 Hz. The task consisted of 205 trials, with 25 of them being 1-back repeats. The third task consisted of a total of 50 stimuli: 10 faces (open-source database[28]), tools (BOSS database[29]), patterns, houses, and body parts. Each image exemplar was presented 4–5 times throughout the task. The task was designed in a block form. Each block consisted of 10 images belonging to the same category, presented in a pseudo-random sequence (with each example displayed once per block, except for 1-back repeats). Images were displayed for 500 milliseconds, followed by a jittered inter-stimulus interval ranging from 750 to 1500 milliseconds. The blocks were separated by either 4 or 8 seconds. The task comprised 260 trials (26 blocks) and 18 1-back repeats.

The patients were seated on a bed in front of an LCD monitor for all three versions of the task. During the tasks, stimuli were centrally presented, with a visual angle of approximately 13° in task 1 and 11° in tasks 2 and 3. Patients were instructed to maintain fixation throughout the task and to click the mouse button whenever they observed consecutive repetitions of the exact same image. For five patients (one in set 1, two in set 2, and two in set 3), the instruction was to press the button on each trial, not just on 1-back repeats, to indicate whether or not a 1-back repeat had occurred.

*iEEG recordings*

Recordings were conducted at North Shore University Hospital, Manhasset, NY, USA. Electrodes were either subdural strips or grids placed directly on the cortical surface (Ad-Tech Medical Instrument, Racine, Wisconsin) and/or depth electrodes (PMT Corporation, Chanhassen, Minnesota). For Ad-Tech, the subdural contacts were 3 mm in diameter and were spaced at 1 cm distances. For PMT, the depth contacts were 2 or 1 mm in diameter and spaced at 2.5 or 5 mm apart. The maximum penetration of depth electrodes was 70 mm, corresponding to a maximum of 13 implanted contacts. The signals were referenced to a vertex screw or a subdermal electrode, filtered electronically (analog bandpass filter with half-power boundaries at 0.07 and 40% of sampling rate), sampled at 512 or 500 Hz, and stored for offline analysis using XLTEK EMU128FS or NeuroLink IP 256 systems (Natus Medical Inc., San Carlos, CA). Electrical pulses were sent upon stimuli onsets and recorded along with the iEEG data for precise alignment of task protocol to neural activity.

*Anatomical Localization of iEEG contacts*

Prior to the electrodes implant, patients were scanned with a T1-weighted 0.8 mm isometric anatomical MRI on a 3 Tesla Signa HDx scanner (GE Healthcare, Chicago, Illinois). Following the implant, a computed tomography (CT) and a T1-weighted anatomical MRI scan on a 1.5 Tesla Signa Excite scanner (GE Healthcare) were collected. The CT taken after implantation was overlaid on top of the post-surgery MRI scan which was then adjusted to the MRI scan taken prior to the implantation surgery using FSL's Flirt. This alignment was then used to visualize the post-implant CT scans as overlayed on the pre-surgical MRI scans. The identification of each contact was then performed manually based on inspection of the CT aligned to the pre-surgery MRI (that is, in the patient's native space) using the BioImage Suite[30].

The localization of iEEG contacts onto the cortical surface was conducted following previous routines[31]. The cortical surface of each patient was reconstructed and segmented based on the pre-surgical MRI scan using the recon-all Freesurfer function (FreeSurfer 5.3). Each contact was then localized to the cortical vertex that was nearest to its estimated position. To project all patients' electrodes contacts onto a single cortical template, the cortical surface was unfolded and for each patient the unfolded spherical mesh of each patient was resampled and projected onto a standard mesh (FSaverage mesh template using SUMA[32]).

*HFA Estimation and Signal Preprocessing*

In this study, we analyzed the high-frequency amplitude (HFA, 48–154 Hz) signal, which has been found to be functionally selective[33,34] and to be a reliable indicator of aggregate firing rate in humans[35,36]. All signals were preprocessed to be 500 Hz, signals originally recorded at a sampling rate of 512 Hz were down-sampled. Raw time-courses and power spectra of all channels were manually inspected, and irregular channels were excluded from further analysis. Next, the signal was divided into nine sub-ranges of 10Hz width to estimate high-frequency amplitude (HFA) modulations, ranging from 48 to 154 Hz. We excluded 59–61 and 117–121 Hz sub-ranges to discard line noise. We performed band-pass filtering on the signal at each frequency sub-range and the instantaneous amplitude in each sub-range was estimated by taking the absolute value of the filtered signal's Hilbert transform[37]. Each sub-range was divided by its mean value, and the normalized values were then averaged across all nine sub-ranges. This was done because the 1/f profile of the signal's power spectrum leads to a higher impact of lower frequencies on the overall estimation of high-frequency activity (HFA). Data preprocessing was carried out using Matlab (R2017a). For the frequency sub-range filtering, we used the original EEGLAB's Hamming windowed FIR filter (pop_eegfiltnew function).

*Contacts inclusion criteria*

Next, we defined visually responsive contacts as we reported in previous studies [38,39]. Briefly, the mean response of each contact to all available stimuli from the versions the patient participated in were compared to baseline (paired t-test on mean exemplar responses versus baseline, at 50–500 ms and −200 to 0 ms relative to image onset, respectively). Hit, miss, and false alarm trials were excluded from all analyses. FDR correction was then applied to the pooled p values from all 68 patients. Contacts with pFDR < 0.05 and a considerable effect size (Glass' Δ) of larger than 1 were defined as visually responsive. Among all contacts implanted in 68 patients, 852 contacts were found to be visually responsive. For our analyses, we next considered only visual contacts that were localized in high-order visual areas, defined anatomically as the conjunction of labels comprising the ventro-temporal cortex (VTC). Anatomical labels were taken from the FreeSurfer parcellation of individual brains in their native space (prior to projection onto the common surface), based on the Destrieux anatomical atlas[40]. Finally, patients with less than 4 high-order visual contacts were excluded from the analyses. This resulted in a total of 19 patients and 140 contacts. Note that due to the within-task nature of our analyses, contacts recorded in two task versions were considered as separate in the analysis, each time with a different functional tunning curve, corresponding to the images of the relevant task. We therefore count 244 contacts from 19 patients across all the 3 sets (see Table S1).

For all subsequent analyses, visual responses were defined as the average HFA response at 50-350 msec following stimulus onset. These amplitudes were averaged across repetitions of the image, however 1-back repetitions and false alarms trials were discarded from the analysis.

*Data Analysis*

*Definitions*

We denote by $a_j^i$ the activation value of contact $j$ when the participant was exposed to image $i$. Accordingly, the term $a_j^*$ denotes the contact values across all images, thus $a_j^*$ represents the *tunning curve* of contact $j$ whose dimensionality equals the number of images (see Figure 1). Having a group of contacts, $\mathcal{A}$, from the same brain (and usually from a similar cortical area), we rearrange the values of these contacts when exposed to image $i$ into an **activation vector** $a_\mathcal{A}^i$ where the dimensionality of $a_\mathcal{A}^i$ equals the number of contacts in the group $\mathcal{A}$. When analyzing the invariant characteristics of activations from multiple brains, we compare all possible pairs of contacts, each of each from a different brain.

*Permutation inter-subject matching*

As typically is the case with intracranial recordings, the number and location of contacts is different among subjects. The first step, referred to as permutation, is thus intended to find a mapping between the contacts of each pair of subjects. Due to the limitations of anatomical distances (see Figure 4), we opt for a cross-subject mapping that optimizes activation correlation.

Assume subjects $A$ and $B$ with contact $\mathcal{A}$ and $\mathcal{B}$, respectively, where $|\mathcal{A}| = n_a$ and $|\mathcal{B}| = n_b$. Without loss of generality, we assume $n_a \geq n_b$. In order to find the optimal contact match between the subjects, we construct a bipartite graph $G = (\underline{\mathcal{A}}, \underline{\mathcal{B}}, E)$. Each node $i \in \underline{\mathcal{A}}$ represents a contact in subject *A*, and each node $j \in \underline{\mathcal{B}}$ represents a contact in subject *B*.

Each edge $e_{i,j} \in E$ from node $i \in \mathcal{A}$ to node $j \in \mathcal{B}$ is weighted by the Spearman correlation between the tunning curves $a_i^*$ and $a_j^*$: $e_{i,j} = S(a_i^*, a_j^*)$. In order to find the best matching between the contacts we use the *max-weight matching* algorithm in a bipartite graph[41] which finds the optimal match between the contacts, such that the total Spearman correlations are maximized. Since $n_a \geq n_b$, the $n_b - n_a$ non-matched contacts in $\mathcal{B}$ are discarded. Having an optimal matching between contacts in $\mathcal{A}$ and $\mathcal{B}$, we rearrange the $n_a$ contacts in $\mathcal{A}$ and in $\mathcal{B}$ in the same order. This means, that for each image $k$, the *activation vectors*, $\boldsymbol{a}_\mathcal{A}^k$ and $\boldsymbol{a}_\mathcal{B}^k$ are both vectors in $\mathbb{R}^{n_a}$ (see Figure 1) and where $\forall i, \boldsymbol{a}_\mathcal{A}^k[i]$ was matched to $\boldsymbol{a}_\mathcal{B}^k[i]$. This procedure is applied to each pair of subjects.

*Relational coding: Inter-subject Rotational Invariance*

The second hypothesis we consider is that of *relational coding*. Formally, a *relational coding scheme* implies that the activation codes are invariant across individuals up to orthogonal transformations, namely, for each pair of subjects, $\mathcal{A}$ and $\mathcal{B}$, there is an orthogonal transformation $R$, such that, for each image $i$, $R\boldsymbol{a}_\mathcal{A}^i \approx \boldsymbol{a}_\mathcal{B}^i$.

First, for each category of visual stimuli and each pair of patients we applied the optimal-match procedure as discussed above. Next, we computed the optimal orthogonal transformation for each pair of patients on the training set, and then applied the evaluated rotation to the validation set. The optimal rotation, $\hat{R}$, was evaluated by solving:

$$\hat{R} = argmin_R \sum_{i \in train\_set} \left\| R\boldsymbol{a}_\mathcal{A}^i - \boldsymbol{a}_\mathcal{B}^i \right\|_2$$

This optimization is known as *Orthogonal Procrustes Problem*[42,43] where the optimization is performed in the following manner: we construct two matrices from the activation vectors:

$$\boldsymbol{a}_\mathcal{A}^* = [\boldsymbol{a}_\mathcal{A}^1, \boldsymbol{a}_\mathcal{A}^2, \cdots, \boldsymbol{a}_\mathcal{A}^n] \text{ and } \boldsymbol{a}_\mathcal{B}^* = [\boldsymbol{a}_\mathcal{B}^1, \boldsymbol{a}_\mathcal{B}^2, \cdots, \boldsymbol{a}_\mathcal{B}^n]$$

It can be shown that the problem is equivalent to finding the nearest orthogonal matrix to the matrix $E = (\boldsymbol{a}_\mathcal{B}^*)(\boldsymbol{a}_\mathcal{A}^*)^T$ which means solving the following problem:

$$\hat{R} = \min_R \|R - E\|_F \quad s.t. \quad R^T R = I$$

The solution can be calculated using the SVD decomposition of $E$: $E = U\Sigma V^T$ and $\hat{R} = UV^T$

**Linear coding: Inter-subject linear Invariance**

Finally, in order to test a more general coding hypothesis, we find the general linear transformation. As in the Relational Coding, we computed the optimal linear transformation for each pair of patients on the training set, and then applied the evaluated transformation to the validation set. The optimal linear transformation, $\widehat{M}$, was evaluated by minimizing:

$$\widehat{M} = argmin_M \sum_{i \in validation\_set} \left\| M\boldsymbol{a}_\mathcal{A}^i - \boldsymbol{a}_\mathcal{B}^i \right\|_2$$

Where M is a general linear transformation matrix. Using the above notations for $a^*_{\mathcal{A}}$ and $a^*_{\mathcal{B}}$, $M$ can be estimated using Moore–Penrose pseudo inverse:

$$\widehat{M} = a^*_{\mathcal{B}} \, (a^*_{\mathcal{A}})^T \, [\, a^*_{\mathcal{A}} (a^*_{\mathcal{A}})^T \,]^{-1}$$

## Acknowledgements:

This work was supported by the Israeli Ministry of Science and Technology under The National Foundation for Applied Science (MIA).

**Supplementary Figures and Tables**

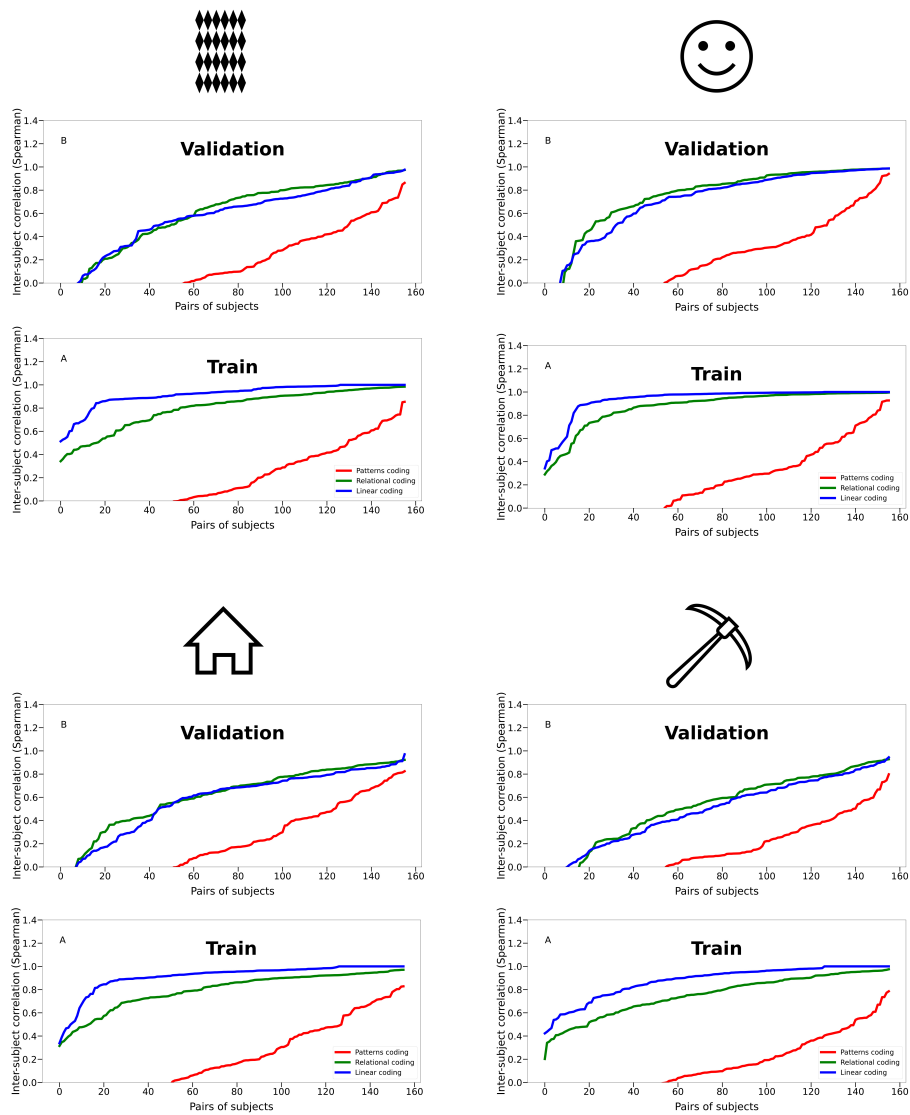

**Figure S1**. Inter-patient ISSI of the three neuronal coding schemes for textures, faces, buildings and tools categories in set1. Similar presentation format as Figure 5 in the main text. The Y-axis depicts the Spearman correlation between pairs of patients. X-axis depicts the pair index. The ISSI values are sorted and shown in increasing order. Bottom part of each panel – training set. Top part of each panel – the independent validation set. Red, green, and blue colors indicate the ISSI for optimal permutation (activation vector matching), rotation (relational code matching), and linear transformations matching, respectively. Note that for the validation set, across all four categories, the relational coding scheme (rotation, green) shows a clear enhancement compared to the other coding schemes.

| Dataset | Number of patients | Number of high-order visual contacts | After minimal contacts filter ||
|---|---|---|---|---|
| | | | Patients | Contacts |
| 1 | 37 | 128 | 13 | 91 |
| 2 | 45 | 173 | 17 | 129 |
| 3 | 13 | 36 | 3 | 24 |

**Table S1**: Patients' and contacts' numbers included in the analyses, per dataset. Patients for which the number of high-order visual contacts was smaller than 4 were excluded from the analyses.